# Overhead Management in Multi-Core Environment

Urmila Shrawankar and Mayuri Joshi
RTM Nagpur University, Nagpur (MS), India
urmila@ieee.org

**Abstract:** — In multi-core systems, various factors like inter-process communication, dependency, resource sharing and scheduling, level of parallelism, synchronization, number of available cores etc. influence the extent of possible High Performance Computing parallelization. These parameters if not managed to the root level, later surface as overheads during execution. This paper emphasizes on these parameters of parallelism, their overheads of parallelization and its effective management for optimal parallel execution under any domain. As a whole, we focus on the Dense Linear Algebra (DLA) domain and specifically on Matrix Multiplication and sorting domains. These domains are chosen as they find application in various sectors of scientific and mathematical applications. The comparative analysis of results obtained clarifies the trade-off between serial and parallel execution of DLA problems the surfacing overheads and their possible and effective management.

### Introduction

When we deal with multicore environment or parallel processing concept, most important thing to be considered is the Amdahl's Law [3] for parallel computing. As an abstract view of this law, we understand that increasing the number of cores can lead to more efficient High Performance Parallel computing [2] scenario. But the changing nature of hardware and its optimality with underlying software platform create an important criticism to the concept of Amdhal's law. This challenge states that, only increasing the number of employed cores cannot optimize the results on parallel systems. The problem of core management becomes more crucial when we are dealing with multi-core systems [12] with shared memory space and limited availability of cores.
Following are the reasons for the above discussed challenge.

1. Dependency [1] prevailing in the working of different cores operating independently and in parallel [13]
2. Size of problem being solved should be comparable to the efforts necessary for dividing the tasks into partitions
3. Overhead of task division when either sub tasks under consideration are not independent enough or the size of problem is too small to be divided and employed on parallel platforms.

This paper mainly emphasizes on the scope of overhead management on multi-core systems for Dense Linear Algebra (DLA). The work is mainly performed on three different domains of DLA problems in terms of their extent of dependency and scope for parallelization.

## Overheads of parallelism in Matrix Multiplication and their Management

For Matrix multiplication or matrix chain multiplication problems, parallelization can be implemented because all input output operations do not depict dependency. Elements for single row column multiplication are interdependent but operations are not dependent. Thus if each row column operation is fragmented, it will create overhead of synchronization at each step of inter product additions. Ultimately leading to the inter core communication overhead.

Our main focus is for the management of these prominent overheads as well as the analysis for the same. Table 1 focuses on the parameters of serial and parallel execution for the Matrix multiplication. It also puts a lime light on the possible overheads and their management.

As discussed in the above work, we understand the actual gap of working scenario of serial and parallel concepts. Figure 1 describes the way out for the overhead occurrence and scope for its management.

In the fig.1, under overhead reasoning we state the possible overhead occurrence while problem scope depicts the pictorial execution of the problem. In the section of methodology for overhead management, the possible ways of implementation are discussed. The technique of fork joining [5], [10], [11] is used for switching between the serial and parallel computations.

Table 1: Comparative scope analysis for parallelization of Matrix multiplication.

| Parameter | Scope of Serialization | Scope of Parallelization |
|---|---|---|
| **Order of matrix** | Best suited for lower order matrices | Best suited for very high order matrices of minimum 1000 and above |
| **Input management** | Being single core execution, complete input is managed by working core only | Input will be dealt with in master slave fashion [10], [11]. The master thread will distribute the row column sets among the available cores. |
| **Processing methodology** | Row column multiplications and inter product addition operations to be carried out in iterative fashion[10] in serial order of occurrence of the rows | All the row column operations are to be distributed among working cores so as to avail the benefit of parallelization |
| **Time requirements** | More time required for very large order matrices due to iterative nature | Time consumed is more for lower order matrices due to overhead of thread creation. For higher order computations, time is saved due to full utility of available cores of system |
| **Nature of overhead** | The large number of row column operations surface like an overhead due to the repetitive nature of common computations again and again | For lower order matrices, overhead of thread creation is seen as well as the need of inter-core communication [7],[8] creates overhead of time management .Synchronization[4] is required for the replication of output matrix |

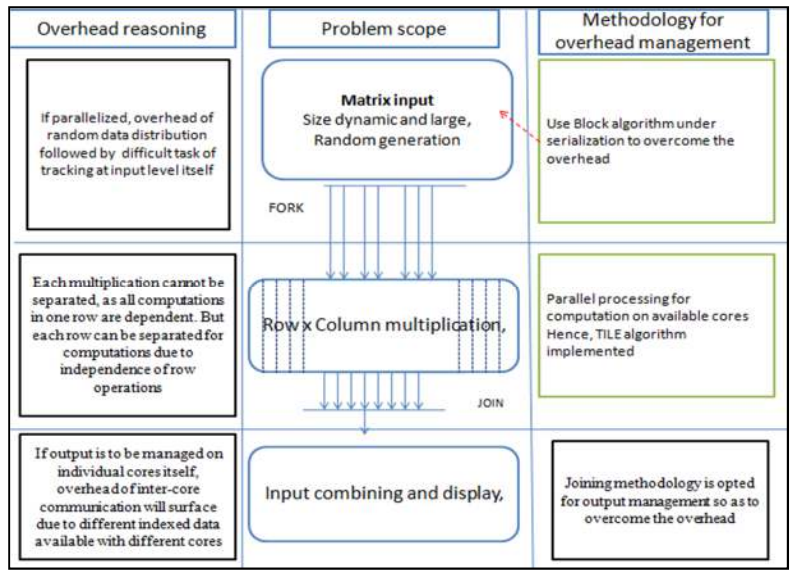

Figure 1: Overhead analysis of matrix multiplication on parallel platforms and scope for its management.

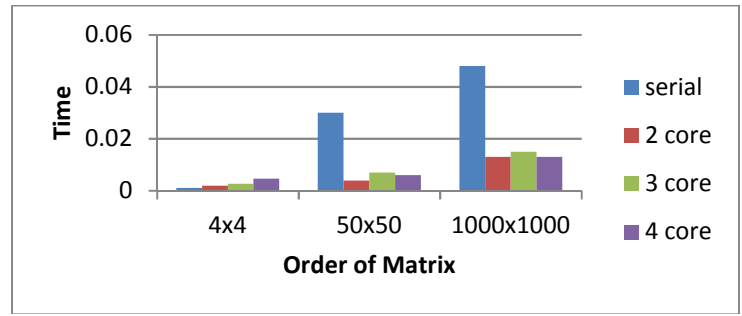

Figure 2: Graphical representation of results for matrix multiplication on serial and parallel platforms.

From the outcomes of figure 2, Graphical representation of results for matrix multiplication on serial and parallel platforms. Parallel implementation is done after overhead management. It is seen that serial execution is very useful for lower order matrices but as order increases, core based execution is seen to be more practical and optimal.

**Overheads of parallelism in sorting and their management**

Under this problem domain, we have taken quick sort for the analysis and study as it is the simple and most popularly parallelized sorting algorithm. Figure 3 shows the basic algorithm for the implementation of quick sort technique. Programs using divide and conquer [6] techniques are not iterative in nature but focus on recursive execution of that part of code to be repeatedly executed. This requirement of recursive approach creates overhead of data distribution for processing on multiple platforms. In fig.3 we can see the highlighted part of recursive call for quick sort function. This is the part of code which shows the scope of parallelization for quick sort algorithm as each independent array will not have inter-dependent elements. Only the elements allocated to each individual core will show internal dependence thus increasing the scope of parallelization.

```
1.   procedure QUICKSORT (A, q, r)
2.   begin
3.      if q < r then
4.         begin
5.            x := A[q];
6.            s := q;
7.            for i := q + 1 to r do
8.               if A[i] ≤ x then
9.                  begin
10.                    s := s + 1;
11.                    swap(A[s], A[i]);
12.                 end if
13.           swap(A[q], A[s]);
14.           QUICKSORT (A, q, s);
15.           QUICKSORT (A, s + 1, r);
16.        end if
17.  end QUICKSORT
```

Figure 3: Algorithm for quick sort serial execution

Table 2 elaborates about the parametric analysis of scope of parallelization for this domain. Minute parameters are considered here for detailed analysis.

Table 2: Parametric analysis for quick sort execution on parallel systems

| Parameters | Analysis for parallelization |
|---|---|
| Dependence | Pivot selection and its final placement |
| Input | Complete array of n numbers, initially managed by master thread. |
| Pivot selection | Random , mean, leftmost element, rightmost element |
| Pivot placement | By master thread for avoiding the overhead of re-analysing the pivot given by each core and then swapping the position. |
| Scope of parallelism | Once the initial placement of the pivot is done, the array before the pivot will be allocated to one core and later to the other.<br>As per this technique, each core will further divide the allocated arrays recursively following the same core allocation |
| Output | Collective data of all system core executions.<br>Not considered in parallel section to avoid multiple copies of same index. |
| Overhead observed | Placement of pivot on each core, if list of numbers distributed initially on available cores.<br>Inter-core communication and thread creation overhead if, number of elements is very less |

Figure 4 details about the work flow of quick sort implementation on parallel platforms. It depicts the overall process of dependency analysis for overhead identification for further processing.

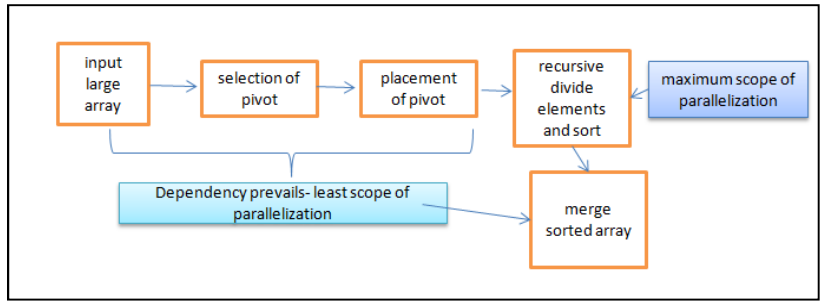

Figure 4: Work flow for execution of quick sort on parallel platform

Based on the detailing and analysis of quick sort implementation on parallel platforms seen in the above section, following results are obtained by various ways of pivot selection as discussed above in table 2. Table 3 gives the detailed results from system execution on Windows platform using OpenMp as the basic for parallelization. Parallel section concept of OpenMp is used as we have to parallelize a complete section of code on available cores.

Table 3: Comparative results of serial to parallel execution of quick sort after removal of possible overheads.

| Elements | serial | parallel left pivot | parallel mean pivot | parallel right pivot | parallel random pivot |
|---|---|---|---|---|---|
| 1000 | 2.246 | 1.4 | 1.247 | 1.37 | 2.293 |
| 1100 | 2.403 | 1.57 | 1.714 | 1.68 | 2.512 |
| 1500 | 3.682 | 1.65 | 1.839 | 1.932 | 2.824 |
| 2000 | 3.838 | 2.074 | 1.933 | 2.151 | 3.136 |

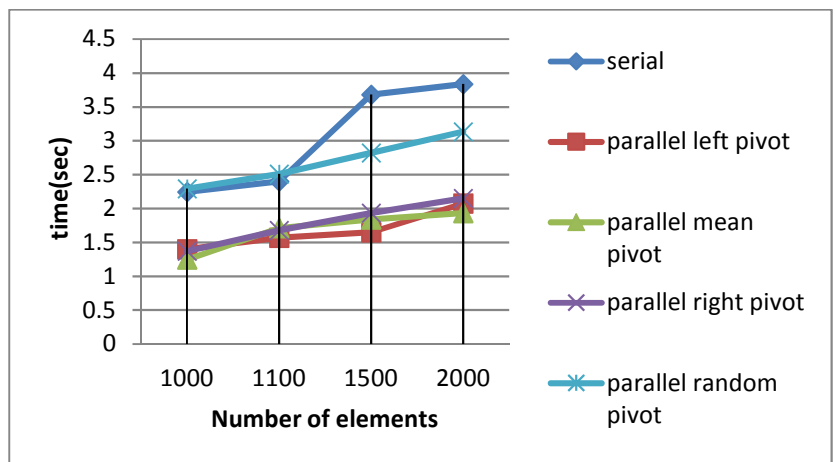

Figure 5: Graphical representation of the results depicted in table 3

From the results of quick sort implementation under different scenarios, we observe that random pivot selection is a domain which requires more analysis. It is observed that time of execution under random selection is exceeding all other methodologies of pivot selection. Also, serial execution here shows comparable results for fewer elements but trade-off increases with the increase in elements.

## Conclusion

Based on the complete work and its analysis we conclude that parallelization is a boon for Dense Linear Algebra domain provided the overheads are identified to the root level and its management is attempted. For the domain of Matrix Multiplications as well as Sorting, parallelization is less effective for lower order inputs. The study and implementation of two different problem scopes clarify that pattern of dependency and overheads identified; vary as per the nature of that problem. Thus each problem space requires detailed and independent analysis of its level of parallelism as well as its scope. We emphasize on this concept of individual analysis due to the fact that parallelization if not implemented properly, will definitely appear as an overhead for execution ruining the speedup of processing. This work finds application in different scientific and mathematical domains where parallelization of mathematical concepts is demanded.